# Absorption of infra-red radiation by atmospheric molecular cluster-ions

K.L. Aplin[*] and R.A. McPheat

Space Science and Technology Department

Rutherford Appleton Laboratory

Chilton, Didcot

Oxon OX11 0QX, UK

**Abstract**

Results from previous laboratory experiments indicate that both the protonated water dimer $H_3O^+(H_2O)$, and molecular cluster-ions, $X^+(H_2O)_n$ absorb infra-red (IR) radiation in the water vapour continuum region between 4-14µm (2500-714cm$^{-1}$). Protonated water clusters are a common species of atmospheric molecular cluster-ion, produced by cosmic rays throughout the troposphere and stratosphere. Under clear-sky conditions or periods of increased atmospheric ionisation, such as solar proton events, the IR absorption by atmospheric ions may affect climate through the radiative balance. Fourier Transform Infrared Spectrometry in a long path cell, of path length 545m has been used to detect IR absorption by corona-generated positive molecular cluster-ions. The column concentration of ions in the laboratory spectroscopy experiment was estimated to be ~10$^{13}$ m$^{-2}$; the column concentration of protonated atmospheric ions estimated using a simple model is $\sim 10^{14}$ m$^{-2}$. Two regions of absorption, at 12.3 and 9.1μm are associated with enhanced ion concentrations. After filtering of the measured spectra to compensate for spurious signals from neutral water vapour and residual carbon dioxide, the strongest absorption region is at 9.5 to 8.8 μm (1050 cm$^{-1}$ to 1140 cm$^{-1}$) with a fractional change in transmissivity of 0.03 ± 0.015, and the absorption at 12.5 to 12.1 μm (800 cm$^{-1}$ to 825 cm$^{-1}$) is 0.015 ± 0.008.

---

[*] Corresponding author: email k.l.aplin@rl.ac.uk

**Keywords**

Atmospheric electricity, atmospheric ions, infra-red radiation, water vapour continuum, natural climate variability

**1. Introduction: atmospheric ionisation**

The electrical conductivity of air is sustained by continual ionisation from cosmic rays. Cosmic rays are highly energetic particles that enter Earth's atmosphere from space. The number of ions created by cosmic rays depends on air's interaction cross-section, which is proportional to its density. Many primary cosmic ray particles pass through the thin ionosphere and mesosphere without encountering any air molecules to ionise, but at a pressure surface of 100-200 hPa, (in the upper troposphere or lower stratosphere), the air density is sufficiently high for the primaries to collide with air molecules (Sandström, 1965). These collisions form a cascade of secondary subatomic particles, which lose energy through causing ionisation events as they propagate downwards. The peak cosmic ray atmospheric ionisation rate occurs at the layer where the secondary particles are produced, and decreases down to the surface, where cosmic ionisation rates are $\sim 2$ cm$^{-3}$s$^{-1}$. Ionisation close to the terrestrial surface is enhanced by the emission of natural radioactive gases, such as radon, from the soil so that the total surface ionisation rate is ~10 $cm^{-3}s^{-1}$. Above the boundary layer, ~1km, the ion concentration increases with altitude.

Ionisation of a gaseous air molecule such as $N_2$ produces a primary air ion e.g. $N_2^+$ and an electron. The electron quickly attaches to another molecule to give a negative primary ion. Primary ions become stable by reacting with other atmospheric molecules, and through charge-driven clustering with polar molecules like water and ammonia (Harrison and Carslaw, 2003). Atmospheric ions are charged molecular clusters, usually $X^+(H_2O)_n$ or $Y^-(H_2O)_n$ (common species are listed in Harrison and Carslaw (2003)). Atmospheric ions are sufficiently mobile to permit electric current flow in response to a potential difference, and form the fair weather conduction part of the global electric circuit. Between the ionosphere and surface, a net conduction current flows of density ~2 $pAm^{-2}$. The electrical conductivity of tropospheric and stratospheric air $\sigma$ is related to the total bipolar ion number concentration $n$, and the mean ionic mobility $\mu$. Typically, atmospheric cluster-ions have $\mu$~1 $cm^2V^{-1}s^{-1}$ and a radius of 0.5 nm (*e.g.* Hõrrak *et al*, 2000). Ion number concentrations $n$ in the atmosphere depend on a balance between the ionisation rate and the number concentration of any aerosol present.

Whilst many aspects of terrestrial atmospheric electricity have been studied quantitatively for over two centuries, understanding of the relevance of atmospheric ionisation to climate through the radiative balance has only recently begun to develop. Discussion of the topic has increased following evidence that cosmic rays could affect climate (Carslaw *et al*, 2003). Connections between cosmic ray ionisation and changes in radiative properties have been suggested on many timescales, from half a billion years (Shaviv, 2002) down to a few minutes (Harrison and Aplin, 2001). One postulated mechanism is that ions formed by cosmic rays enhance the abundance of particles that may act as condensation nuclei for cloud formation. This ion-induced particle formation is plausible in Earth's atmosphere (e.g. Vohra *et al*, 1984), but the quantitative links between it and cloud formation are not yet firmly established. Another mechanism is that the potential of the ionosphere could be changed by long-term variations in cosmic rays, modifying the conduction current and charges on clouds (Harrison and Carslaw, 2003). However, making the necessary measurements to confirm or refute both these effects is difficult, and the hypotheses are primarily supported by models (Lovejoy *et al*, 2004; Yu and Turco, 2001; Tripathi and Harrison, 2001).

This paper investigates a further mechanism by which cosmic ray ionisation could modulate radiative processes: the infra-red (IR) absorption of atmospheric ions. Following laboratory studies by Carlon (1982) in which artificially produced charged clusters absorbed IR radiation in a long path cell, Aplin (2003) suggested that this process should be expected to exist in the atmosphere. Direct absorption of IR radiation by ions is another potential mechanism by which heliospheric changes could affect Earth's radiative balance through modulation of cosmic ray ion production. It is likely that this effect is usually dominated by more significant changes, for example IR emission from clouds. However it is well known that under some conditions atmospheric ionisation can increase by orders of magnitude, and in these cases IR absorption could become more significant. For example during solar proton events, energetic solar particles can produce $10^{32}$ atmospheric ions at ionisation rates of 600-800 $cm^{-3}s^{-1}$ (Bazilevskaya *et al*, 2000). If, as occasionally happens, these solar particles penetrate downwards through the entire atmospheric column to the surface, there is a greater chance of a detectable radiative effect. This presents a further motivation to determine the magnitude of charged cluster absorption. Another reason is to permit accurate assessment of the neutral cluster contribution to the water vapour continuum absorption, which is still poorly understood.

This paper describes laboratory experiments designed to identify and quantify the presence of IR absorption by ion clusters typical of those present in the atmosphere. Evidence for the atmospheric IR absorption of cluster-ions, and suggested mechanisms are discussed in section 2. Laboratory spectroscopy experiments using artificially enhanced ion concentrations to increase the IR absorption of charged clusters are described, and the results and data analysis presented in section 3. Finally, the significance of the findings is assessed in section 4.

**2. Infra-red absorption by cluster ions**

The IR atmospheric radiative properties of the water molecule result from rotational and vibrational transitions of its molecular bonds (*e.g.* Houghton, 2002). There also exists a poorly understood weak IR absorption in the 8-50 $\mu$m (200-1250 cm$^{-1}$) region, with pressure and temperature dependency greater than any known water vapour absorption. This anomaly between accepted theory and observations is referred to as the IR continuum problem. The possible presence of atmospheric water dimers, $(H_2O)_2$, is a long-established explanation for the properties of the IR continuum region (Bignell, 1970). Recent theoretical work implies that atmospheric clusters, particularly dimers and trimers, exist in sufficient concentrations to account for the continuum (Goldman *et al*, 2001; Evans and Vaida, 2000).

**2.1 Laboratory experiments**

Asmis *et al* (2003) confirmed the IR signature of charged water clusters by observing protonated water dimer $H_3O^+(H_2O)$ absorption at 6-14 $\mu$m (1666-714 cm$^{-1}$). The absorption mechanism is by hydrogen bond transitions, similar to those responsible for the IR absorption of neutral clusters. Carlon (1982) used a radioactive source to generate charged water clusters in a humid atmosphere and demonstrated that this mixture of species showed IR absorption bands between 4-13 $\mu$m (2500-746 cm$^{-1}$) over path lengths of 50-100m. Ion concentrations were ~$10^4$-$10^6$ cm$^{-3}$ giving column concentrations of ~$10^{11}$-$10^{14}$ m$^{-2}$. Carlon and Harden (1980) hypothesised that the attachment and recombination reactions between the different clusters caused the IR absorption signal.

**2.2 Absorption by atmospheric ions**

As was discussed in Section 1, hydrated molecular clusters are formed in the atmosphere by cosmic ray ionisation. The protonated water cluster $H_3O^+(H_2O)_n$ is a common atmospheric ion species, and many positive and negative ion clusters are partially or completely hydrated. The existence of a size distribution of atmospheric ions is well established, through both mass spectrometry (Eisele, 1988), and ion mobility spectra (Hõrrak *et al*, 2000). Attachment and recombination reactions between charged atmospheric species are also well known (e.g. Harrison and Carslaw, 2003), and contribute to the variability of the ion mobility spectrum. Therefore it appears reasonable to expect at least some atmospheric ion species to absorb in the IR continuum region.

As mentioned in Section 1, the atmospheric ion concentration varies with height because of changing ionisation rates throughout the atmosphere. Near the surface, radioactive emissions from the ground dominate, so the ionisation rate decreases slightly within the boundary layer. Above about 1 km the cosmic ray ionisation rate increases with height up to ~30 km, after which it is limited by decreasing atmospheric density. The ionisation rate parameterisation of Makino and Ogawa (1985) has been used to estimate the positive ion column concentration $n_+$, for which there is most evidence for IR absorption (Asmis *et al*, 2003), in an aerosol-free atmosphere up to 50 km. The calculation assumes that ion concentrations are only limited by self-recombination with coefficient $\alpha$ so the positive ion concentration $n_+ = \frac{1}{2}\sqrt{\left(q/\alpha\right)}$ where $q$ is the ionisation rate (see Eq 2 in Section 3.2). The assumption that self-recombination dominates over attachment to aerosol particles is restricted to the troposphere and above, as atmospheric aerosol concentrations are negligible outside the boundary layer. Ionisation rate $q$ depends on geomagnetic latitude, which modulates cosmic ray penetration into the atmosphere. In this example a geomagnetic latitude of 50º has been assumed to give an average estimate for ion concentration, which is highest near the geomagnetic poles and lowest at the geomagnetic equator. Typical ion pair production rates are shown in Figure 1a, with the estimated positive ion profile in Figure 1b. Integrating over 50 km to obtain an ion column concentration, as shown in Figure 1c, gives a column concentration of ~$10^{14}$ m$^{-2}$, which is proportional to the positive electrical conductivity of air in the column.

The vertical column concentration of dimers, which are thought to account for much of the water vapour continuum is ~$10^{22}$-$10^{23}$ m$^{-2}$ (Vaida *et al*, 2001). As this is six orders of magnitude greater than the estimated column abundance of ions, it is almost certain that any IR absorption from ions would only contribute a small fraction of the water vapour continuum absorption under the standard conditions represented by Figure 1. However, the distribution of atmospheric ion species is not well known, particularly in the free troposphere, and may be very sensitive to atmospheric trace gas concentrations, aerosol concentrations and ionisation rates. Furthermore, dramatic increases in ion concentrations can be expected during periods of strong solar activity. For these reasons, laboratory experiments have been developed to measure ionic IR absorption at high ionisation rates with the aim of estimating the magnitude of the atmospheric signal. Laboratory experiments are not completely representative of the many complex atmospheric feedback processes and cannot conclusively prove the existence of an atmospheric effect. However, results from experiments in controlled conditions can provide valuable constraints for subsequent atmospheric experiments by estimating magnitudes and sensitivities over a range of conditions.

## 3. Molecular spectroscopy experiments

### 3.1 Methodology

The spectroscopy experiments in this section were all carried out at the UK Natural Environment Research Council (NERC) Molecular Spectroscopy Facility (MSF), located at the Rutherford Appleton Laboratory (RAL). The MSF long path-length absorption cell (LPAC) at RAL (Ballard *et al*, 1992) is a 9-metre long stainless-steel vessel containing multi-pass reflective optics (White, 1942) for broadband high-resolution spectroscopy at long optical path-lengths of up to ~1 km. As the maximum LPAC path length is much shorter than vertical tropospheric path lengths, any integrated signal from natural surface ion concentrations would be smaller than an integrated atmospheric signal from a vertical column. A corona ion source was used to artificially enhance the ionisation rate $q$ and compensate for the short path length. Positive corona was selected, as the strongest evidence for IR absorption is for positive ions (Asmis *et al*, 2003; Carlon, 1982). A Programmable Ion Mobility Spectrometer (PIMS) (Aplin and Harrison 2000, 2001) was used to measure ion concentrations inside the LPAC. Using artificial air (to be described in section 3.3), the LPAC temperature and humidity can be kept relatively constant, and IR absorption spectra can be compared with the corona source activated and disabled. The relative locations of PIMS, ion source and environmental sensors in the cell are shown in Figure 2.

## 3.2 Ion generation and detection

The PIMS instrument is fully computer controlled and incorporates compensation for temperature-dependent leakage currents (Aplin and Harrison, 2001). For this application, the PIMS sensing capacitor was inserted inside the LPAC, with air sucked out of the LPAC into the PIMS. A flexible metallic pipe was attached to the exhaust of the PIMS to permit recirculation of air back to the LPAC. The LPAC is constructed of ~10mm thick stainless steel, and screens out much of the ionisation from natural radioactivity. For this reason, the ambient positive conductivity of air in the cell (proportional to the number of positive ions; see Eq 1 below), is 3-6 $fSm^{-1}$, lower than typical urban outdoor values of ~10 $fSm^{-1}$ (Aplin and Harrison 2000, 2001).

The positive ion corona source used an Applied Kilovolts HP10P 10kV 1mA positive high voltage supply, connected to 8 sharp stainless steel electrodes of tip diameter ~1mm to geometrically generate the high electric fields required for corona generation. A 24V 80mm dc fan was mounted above the corona electrodes to circulate the ions generated throughout the spectroscopy cell; it had no other effects on the ion concentrations measured during the experiment. The corona source current was monitored, and the voltage applied to the electrodes could also be controlled. Corona onset was at ~3.6 kV for the duration of this experiment. The corona source current started to increase non-linearly and the positive ions measured also increased immediately, once the corona onset voltage was reached, as shown in Figure 3a. This is caused by local breakdown of the air producing positive air ions and electrons. The electrons immediately attach to other air molecules to form negative air ions, which recombine with the positive corona electrode resulting in unipolar positive ion production. (The positive ions are then stabilised by clustering with water molecules, as described in Section 1.) Only the generation of corona ions could be responsible for the synchronous non-linear increases in both positive ion concentration and corona current, as the two measurements were independent. This effect was found to occur at a repeatable value of corona voltage.

The PIMS instrument was originally optimised for natural outdoor ion concentrations, and its femtoammeter was saturated by the much greater concentrations of artificially produced corona ions. Rather than reduce sensitivity at ambient ion concentrations by modifying the PIMS femtoammeter, a Keithley 487 picoammeter was used to detect the ions at the PIMS sensing electrodes, instead of the femtoammeter included in the PIMS sensing electronics (Harrison and Aplin, 2000). Positive air conductivity calculated from measurements using the Keithley picoammeter, with and without the corona source, is shown in Figure 3b. Ion measurements were sampled at 1Hz and logged using the RS232 interface (PIMS femtoammeter) or the GPIB IEEE-4888 interface (Keithley 487 picoammeter).

The ion measurements can be verified by comparing ion production rates estimated using two independent techniques: (1) from the measured conductivity and ion-aerosol theory, and (2) from the current leaving the corona source. The positive ion concentration can be estimated from the positive conductivity and mobility from

$$\sigma_+ \approx n_+ e \mu_+.  \qquad \text{Eq 1}$$

The minimum mobility of ion entering the PIMS sensing capacitor is a function of the applied field, sensor geometry and air flow rate (Aplin and Harrison, 2000), and was calculated to be 0.58 cm$^{-2}$V$^{-1}$s$^{-1}$. If inserted into Eq 1, this can be used to estimate an upper limit on the ion concentration of $10^6$ cm$^{-3}$. In aerosol-free air the steady state ion concentration $n$ is related to the ionisation rate $q$ by a recombination coefficient $\alpha$ (1.6 x 10$^{-6}$ cm$^3$s$^{-1}$) (*e.g.* Harrison and Carslaw, 2003)

$$q = \alpha n^2 . \qquad \text{Eq 2}$$

If the air in the tank is assumed to be free of aerosol particles, then Eq 2 gives the ion production rate as ~1.6 x 10$^6$ cm$^{-3}$s$^{-1}$. The current from the corona source can be used to independently estimate an ion production rate by calculating the number of electrons required to produce the measured corona current (2.25 µA), divided by the volume of the tank (~ 4 m$^3$). This order of magnitude estimate suggested a corona ion production rate of ~2.5 x 10$^6$ cm$^{-3}$s$^{-1}$, which is within a factor of 2 of the ion production rate estimated from the measured ion concentration. It also indicates that the corona source is <100% efficient at ion production, which is expected from leakage currents within the corona source circuitry, or because not all the corona ions generated are detected due to, for example, losses to the LPAC walls. It is conservative to assume that the ion concentration in the LPAC can be measured to ±50%.

If a constant corona source voltage was used, then both the corona currents and the measured ion concentrations were repeatable, Figure 3b. The spectroscopy experiments were carried out at a corona current of 2.25 µA, corresponding to ~$10^6$ ions cm$^{-3}$. This was close to the threshold current for corona emission for this source. Whilst using higher currents would have increased the signal, they were not used in this experiment for two reasons: firstly because of the increased likelihood of trace gases being produced by the corona (this will be discussed in more detail in section 3.3), and secondly to reduce any capacitative coupling effects from the very high electric fields generated from the clouds of space charge inside the LPAC.

**3.3 Spectroscopy**

The spectra were measured on a Bruker IFS 120HR Fourier transform infrared (FTIR) spectrometer at a resolution, defined as (0.9/maximum optical path difference), of 0.03cm$^{-1}$ (0.3nm at 10 μm) over a spectral range of 500-4000cm$^{-1}$ (2.5-20 μm). The spectrometer employed a silicon carbide globar source operating at approximately 1000K, a germanium/potassium bromide (Ge/KBr) beamsplitter and a broadband mercury-cadmium-telluride (MCT) detector, optimal for this wavelength range. For each spectrum 200 scans were co-added taking just over 1 hour.

The optical system of the LPAC provides very long absorption paths through the gas, or mixture of gases, contained in the inner vessel. It comprises three spherical mirrors of 305 mm diameter and 8.000 m radius-of-curvature. The configuration is a modification of the simple 'White' multi-pass optics design (White, 1942), described by Bernstein and Herzberg (1948), in which a side extension (tab) on the field mirror allows the number of non-overlapping images that can be stacked across the field mirror to be doubled, hence doubling the optical pathlength. Each mirror is mounted on kinematic adjusters that allow tilt and focus adjustments to be made from outside the LPAC when the inner vessel is evacuated or filled with gas, and at any temperature. Thus the absorption path length can be adjusted from the minimum of 32.75 m to over 1 km, in steps of 32.00 m. The longest optical pathlength achievable depends on the mirror reflectivity for the wavelengths of interest. For this work gold coated mirrors providing broadband, high reflectivity (>98%) in the infrared spectral region allowed a pathlength of 544.75m. The LPAC was fitted with KBr windows to allow the spectrometer to be connected to the LPAC using a fully evacuated optical path to avoid absorption by atmospheric water vapour and carbon dioxide. Ionic column concentrations in the cell were ~$10^{13}$ m$^{-2}$.

As the PIMS was designed for use at ambient air pressure, the LPAC was filled with synthetic air by pumping with a throttled rotary pump from the centre of the cell whilst maintaining a flow of 79% nitrogen / 21% oxygen mix (Air products Zero Air) over a water bath to humidify the artificial air, into one end of the LPAC. Unfortunately a small amount of natural air remained, which caused some residual carbon dioxide absorption. The relative humidity and temperature of the gas mix inside the LPAC were measured using humidity and temperature sensors (Vaisala Type HMP234) mounted inside the LPAC and in direct contact with the gas, one at each end. The pressures of gas samples contained in the LPAC were measured using a high precision Baratron capacitance gauge system with a 1330 hPa full-scale sensor (MKS Type 690). The humidity and pressure were recorded at 10s intervals via a RS232 interface. The temperatures of gas samples contained in the LPAC were measured using 15 platinum resistance thermometers (PRTs) in direct thermal contact with the gases, and logged using a National Instruments data logger NI-4351 at 10s intervals.

Due to the weak absorption expected a large number of spectra were required, both with the natural background ionisation and the enhancement provided by corona. Averages could then be computed to exclude the effects of natural variability in, for example, trace gases affecting ion composition. To minimise the effects of any thermal or temporal drift in the spectrometer, spectra were recorded alternately with the corona on or off, allowing 30 minutes between spectra for the ion count to stabilise. To increase the amount of data available, spectra were also recorded overnight, with the corona source switched on and switched off on alternate evenings.

In order to eliminate contamination by species created by the positive corona discharge the spectra were compared to simulated spectra for ozone and oxides of nitrogen created using the Hitran database (Rothman *et al*, 2003). No absorption lines of any contaminant were seen. A typical raw spectrum, showing absorption from the residual carbon dioxide, can be seen in Figure 4. The ratio of all the spectra measured with enhanced ion concentrations, compared to spectra measured under ambient ion concentrations showed two absorption bands at 12.3 and 9.1μm (810 and 1095 cm$^{-1}$). These bands are in similar locations to ionic absorption bands identified by Carlon (1982) at 11.8 and 9.3 μm (847 and 1075 cm$^{-1}$), but further analysis was necessary to confirm the magnitude of any ionic absorption. The dominant absorption signals from residual gases such as $CO_2$ prevented unambiguous identification of any other absorption bands.

**3.4 Analysis of spectroscopic data**

To detect continuum absorption due to charged clusters in a spectral region dominated by neutral water absorption lines, it was decided to fit the known neutral water vapour absorption lines and ratio the spectra to obtain just the continuum spectrum. A similar approach was used to compensate for the absorption by residual carbon dioxide. A sinusoidal absorption feature known as channelling, caused by interference between IR radiation being multiply reflected inside one of the optical windows, was evident in the data (see the vertical lines in Figure 5) and was removed by preprocessing. Simulated transmission spectra of water and carbon dioxide were generated using the Reference Forward Model (Dudhia, 1997) with Hitran-2000 data (Rothman *et al*, 2003). A polynomial fit to the background was used to create a simulated absorption spectrum in the 10.20-11.05 µm (905-980 cm$^{-1}$) region to allow the water vapour concentration to be fitted to minimise the residual in this region. The carbon dioxide was fitted in a similar way between 13.04-13.07 µm (756-767 cm$^{-1}$). Using these fitted concentrations of water vapour and carbon dioxide a final simulated spectrum was generated and used to ratio the measured spectra and produce an absorption line free spectrum. Data were excluded where strong absorption lines reduced the transmission to below 25% to avoid areas of saturated absorption in the detector. As this removed sharp features in the spectrum, the resolution was reduced to increase the signal to noise ratio. The spectra for both background and enhanced ionisation were averaged and smoothed using a 9-point Savitzky-Golay filter. The enhanced spectrum was ratioed by the background to remove the spectral response of the spectrometer and is shown in Figure 6. Two regions of enhanced absorption can be seen: from 9.5 to 8.8 µm (1050 cm$^{-1}$ to 1140 cm$^{-1}$) there is a fractional change in the absorption of ~0.03 ± 0.015 and a weaker absorption of ~0.015 ± 0.008 at 12.5 to 12.1 µm (800 cm$^{-1}$ to 825 cm$^{-1}$). The fractional error in these absorption signals is conservatively estimated based on the variability in the corona ion measurements, discussed in section 3.2.

## 4. Conclusions

Figure 6 shows two IR absorption bands measured in the presence of cluster-ion column concentrations of $\sim 10^{13}$ m$^{-2}$. The location of these bands is very close to two of the bands detected by Carlon (1982) in a similar experiment, with ion column concentrations of $10^{11}$-$10^{14}$ m$^{-2}$. It appears likely that the two regions identified at 12.3 and 9.1μm are related to the presence of artificially generated positive molecular-cluster ions causing a fractional change in the IR transmissivity in the spectroscopy cell. As was discussed in section 2.2, typical columnar concentrations of positive atmospheric ions $\sim 10^{14}$ m$^{-2}$. Although atmospheric conditions clearly differ from the LPAC with substantial pressure and temperature variations over a 50km vertical atmospheric column, a similar order of magnitude of absorption could be expected from typical atmospheric ion concentrations. This may be detectable in the atmospheric downwelling longwave radiation at 12.3 and 9.1μm under cloud-free conditions. Further work is required to quantify the sensitivity of the absorption to variations in the ion concentration, and the effect of changing the neutral water vapour concentration.

**Acknowledgements**

This research was funded by the UK Natural Environment Research Council under their New Investigators' Scheme, and by providing access to the Molecular Spectroscopy Facility. We acknowledge technical assistance from R.G. Williams, J.G. Firth, Dr K.M. Smith (RAL), H.C. Brown (Imperial College, London), and helpful discussions with W. Ingram (Hadley Centre).

**Figure Captions**

Figure 1 a) Estimated ion pair production rate variation with height b) Positive ion concentration variation with height c) Integrated positive (+ve) ion column concentration with height. Data in the plots is calculated from the parameterisation by Makino and Ogawa (1985) for an aerosol-free troposphere and stratosphere at a geomagnetic latitude of 50º, in the middle of the solar cycle.

Figure 2 The MSF Long Path Cell, showing approximate beamline position, Fourier Transform Infrared Spectrometer (FTIR) and locations of humidity sensors, corona source and ion counter. Flushing with artificial air ventilates the cell. The corona source includes a fan to distribute ions evenly throughout the cell, and the PIMS instrument also sucks air through its sensing electrode with a fan to ensure a constant supply of ions. (The array of platinum resistance thermometers and pressure sensors along the cell walls is not shown)

Figure 3 a) Positive air conductivity and corona source supply voltage, showing corona onset. b) Positive conductivity, measured with the corona source cycling. Average conductivity with enhanced ion concentrations (corona activated) was ~180fSm$^{-1}$ at a corona current of 2.25µA.

Figure 4 Example raw spectrum, showing absorption by residual $CO_2$ in the region centred around 4.3 and 14.9 μm (2340 and 670 cm$^{-1}$). The region in which absorption from cluster-ions was found is shaded in grey.

Figure 5 Raw spectrum with enhanced ionisation divided by spectrum from ambient background ionisation, showing areas of enhanced absorption at 12.3 and 9.1μm (810 and 1095 cm$^{-1}$). The absorption at 13μm is due to $CO_2$.

Figure 6 IR spectrum at enhanced ionisation levels, filtered to remove the absorption from neutral water clusters and residual carbon dioxide vapour, divided by a similarly filtered ambient ionisation spectrum. Absorption bands, likely to be from molecular cluster-ions can be seen at 12.3 and 9.2 μm (815 and 1090 cm$^{-1}$).

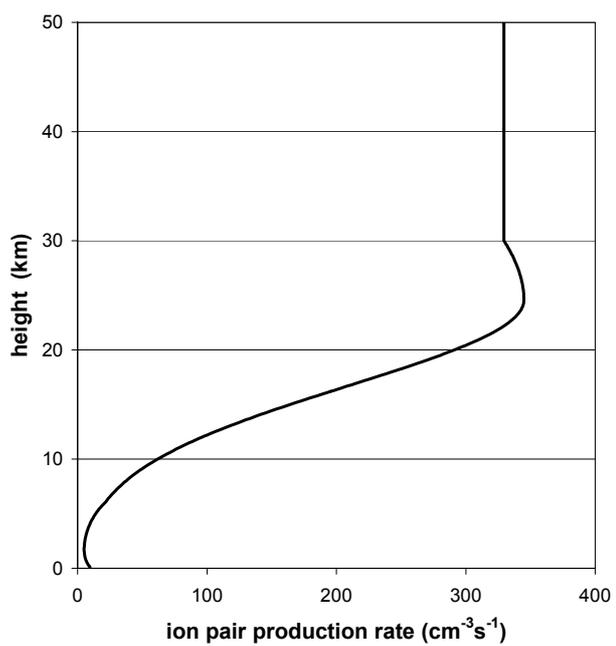

Figure 1a

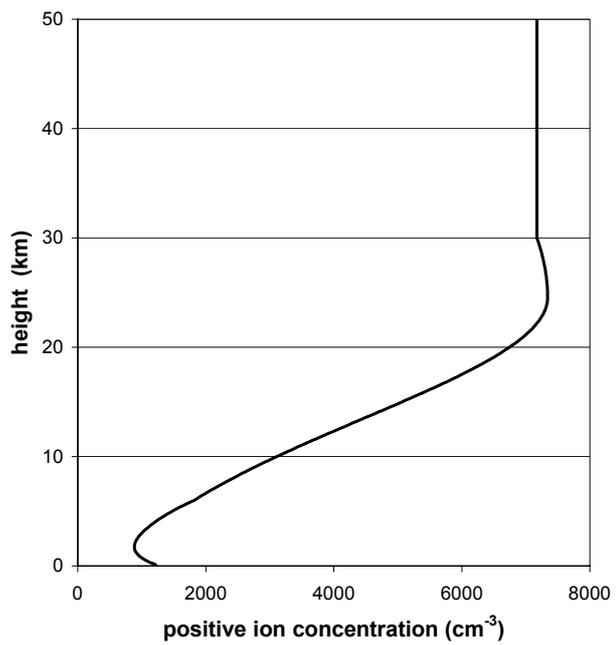

Figure 1b

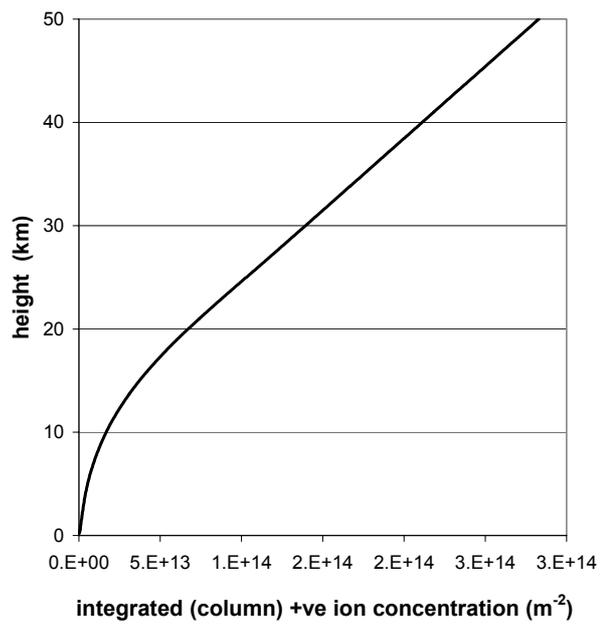

Figure 1c

Figure 2

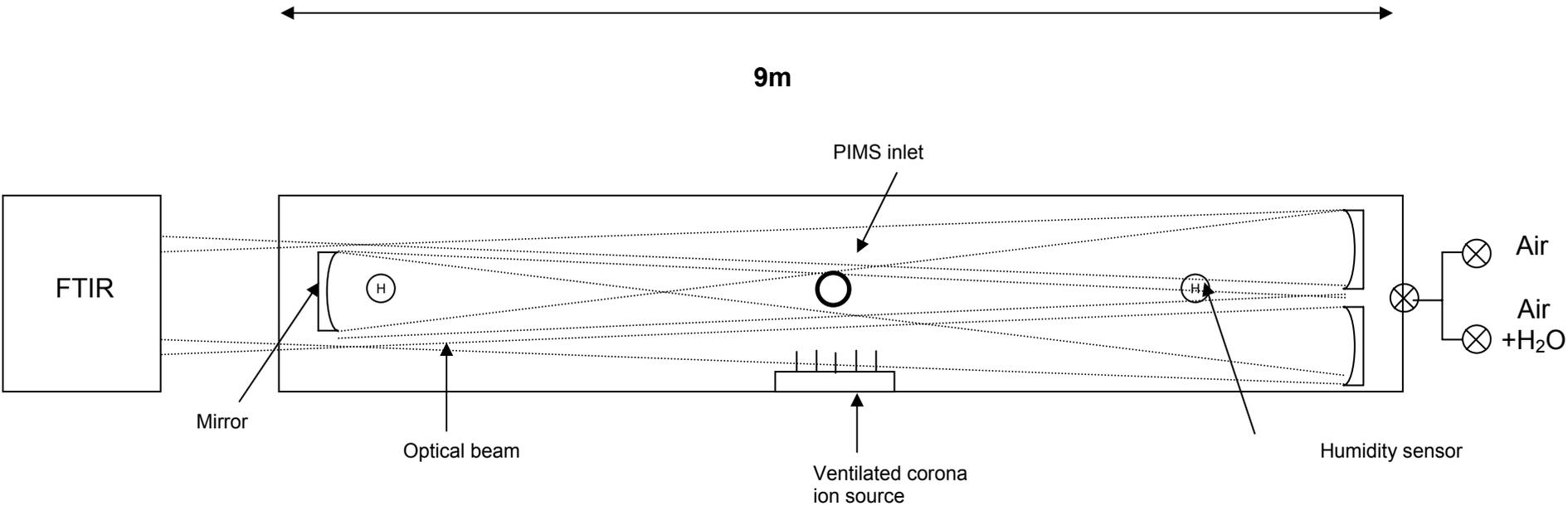

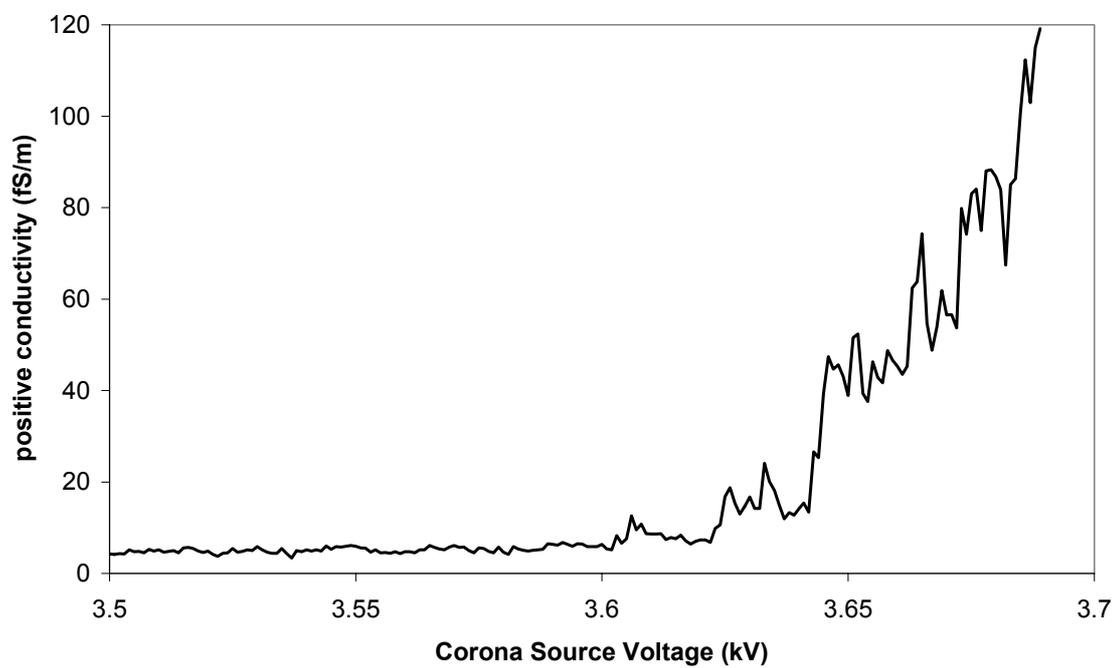

Figure 3a

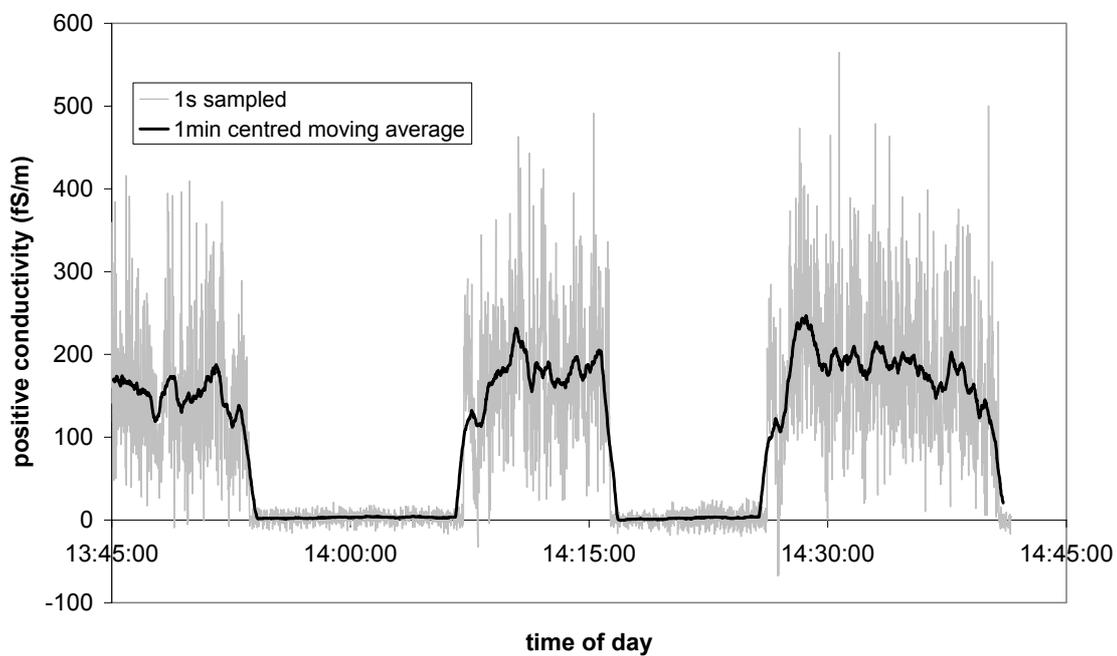

Figure 3b

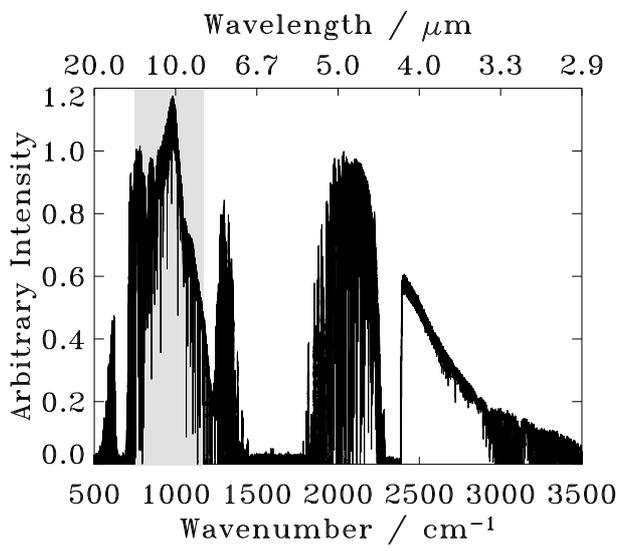

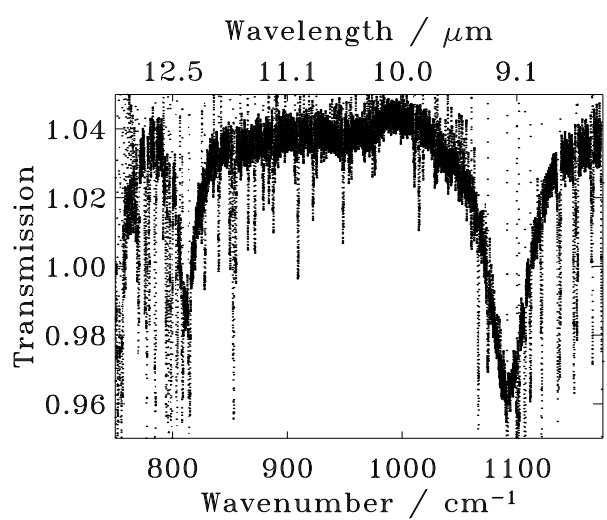

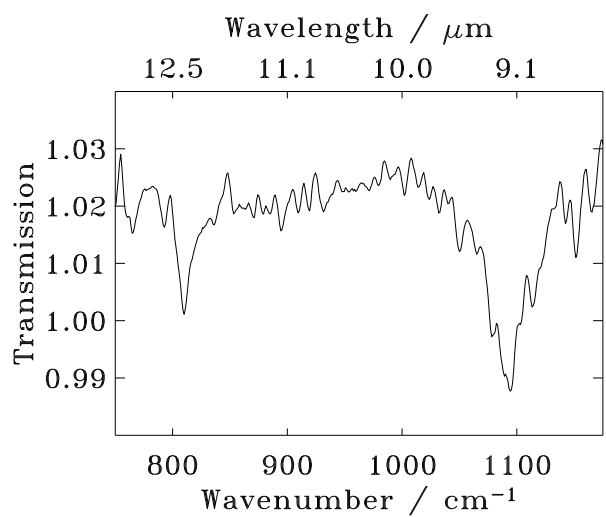